\documentstyle[prl,aps,multicol]{revtex}
\input{epsf}

\begin{document}

\title{Dynamics of a faceted nematic-smectic B front in thin-sample 
directional solidification} 
\author{T. B\"orzs\"onyi$^{1,2}$, S. Akamatsu$^1$ and G. Faivre$^1$} 
\address{$^1$ Groupe de Physique des 
Solides, CNRS UMR 75-88, Universit\'es Denis Diderot and Pierre et Marie Curie,
Tour 23, 2 place Jussieu, 75251 Paris Cedex 05, France\\
$^2$Research Institute for Solid State Physics and Optics,
  Hungarian Academy of Sciences,  H-1525 Budapest, P.O.B.49, Hungary
}
\date{\today}
\maketitle

\begin{abstract}

We present an experimental study of the directional-solidification 
patterns of a nematic - smectic B front.  The chosen system is 
$\rm{C_4H_9-(C_6H_{10})_2CN}$ (in short, CCH4) in $12 \mu m$-thick 
samples, and in the planar configuration (director parallel to the 
plane of the sample).  The nematic - smectic B interface presents a 
facet in one direction --the direction parallel to the smectic 
layers-- and is otherwise rough, and devoid of forbidden directions.  
We measure the Mullins-Sekerka instability threshold and establish 
the morphology diagram of the system as a function of the 
solidification rate $V$ and the angle $\theta_{0}$ between the facet 
and the isotherms.  We focus on the phenomena occurring immediately 
above the instability threshold when $\theta_{0}$ is neither very 
small nor close to $90^{o}$.  Under these conditions we observe 
drifting shallow cells and a new type of solitary wave, called 
"faceton", which consists essentially of an isolated macroscopic facet 
traveling laterally at such a velocity that its growth rate with 
respect to the liquid is small.  Facetons may propagate either in a 
stationary, or an oscillatory way.  The detailed study of their 
dynamics casts light on the microscopic growth mechanisms of the facets 
in this system.

\end{abstract}

\begin{multicols}{2}
\narrowtext
\section{Introduction}
\label{intro}

A crystal growing from an undercooled melt rejects heat and chemical 
species, which must diffuse away in the liquid for the process to 
continue.  The thus generated thermal and solutal gradients tend to 
destabilize the advancing solid-liquid interface.  This effect is 
counterbalanced by the surface tension and the so-called interfacial 
kinetics, which tends to slow down the progression of the interface, 
and hence stabilize it.  As a result of the competition between these 
conflicting factors, solidification fronts may assume a large variety 
of nonlinear patterns, the characteristics of which depend on  the 
control parameters, and the initial and boundary conditions of the process.

The study of solidification patterns has been an active field of 
research for several decades \cite{go92,crho93,la80}.  Most of the 
existing studies are devoted to fully non-faceted systems.  In such 
systems, the surface tension $\gamma$ and the kinetic coefficient 
$\beta$ (defined as the ratio of the kinetic undercooling to the 
growth velocity) are non-singular functions of the orientation of 
the interface with respect to the crystal lattice.  On a molecular 
scale, this corresponds to the fact that the interface is rough in all 
orientations.  Familiar aspects of the dynamics of fully non-faceted 
systems in directional solidification (i.e., when the system is pulled 
at a constant velocity $V$ toward the cold side of an applied 
unidirectional thermal gradient G; see Fig.  \ref{dirsol1}) are the 
existence of a stable planar front at low values of $V$, the primary 
cellular (or Mullins-Sekerka) instability occurring at a threshold 
velocity $V_c$, the quasi periodic arrays of rounded cells at $V$ 
slightly above $V_c$, and of dendrites at $V$ much higher than $V_c$.  
Many dynamical features of these patterns (e.g., stability limits, 
modes of instability) are not yet fully understood, but some of their 
fundamental properties are now clear, among which the crucial role 
played by interfacial anisotropy \cite{go92,cose76,koka96}.  In fact, 
a certain minimum degree of interfacial anisotropy is a necessary 
condition for cellular and dendritic arrays to be stable, or even 
exist.  In thin samples --i.e., quasi bidimensional (2D) systems-- , 
$\gamma$ and $\beta$ are functions of a single variable, say, the tilt 
angle $\theta$ of the interface with respect to the isotherms.  
The functions $\gamma(\theta)$ and $\beta(\theta)$, and thus the solidification 
patterns, depend on the orientation of the crystal with respect to the 
solidification setup \cite{meos90,akfa95,akfa98}.



\vspace*{-0.2cm}
\begin{minipage}{8.3cm}
\begin{figure*}
\epsfxsize=9cm
\centerline{\epsfbox{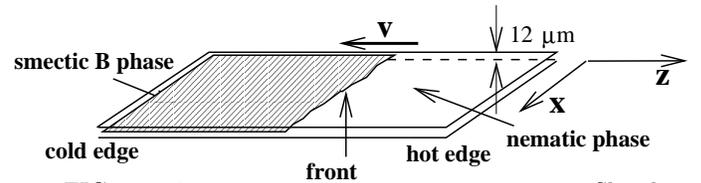}}
\caption{ Sketch of a thin-sample directional-solidification setup.  $z$: 
axis of the thermal gradient.  $x$: axis parallel to the isotherms.  $V$: 
pulling velocity. After a transient, the front advances (in 
average) at the imposed velocity $V$ with respect to the liquid, and 
thus remains essentially immobile in the laboratory reference frame.  
It can then be continuously observed with an optical microscope. } 
\label{dirsol1}
\end{figure*}
\end{minipage}
\smallskip

In contrast with the case of fully non-faceted systems, little is yet 
known about the directional-solidification dynamics of faceted 
crystals.  The few existing experimental studies on 
this subject first of all show that a distinction must be made between 
fully and partly faceted systems \cite{meos90,shhu91,fatr97,osme89,meos91}.
Growth facets (which most 
generally, although not necessarily, coincide with equilibrium facets 
\cite{buca51}) correspond to planes of the crystal containing several 
directions of strong binding.  Fully faceted crystals have numerous 
facet directions, and their directional-solidification fronts consist 
of a succession of facets limited by sharp edges.  The dynamics of 
such fronts does not give rise to any stationary state, in general, 
and bears no obvious relation with that of non-faceted fronts.
Partly faceted systems only have a few facet directions 
connected to one another by large rounded regions.  In lamellar 
crystals, the solid-liquid interface may be rough in all but one 
direction, namely, that of the molecular layers. In this case, when the tilt 
angle $\theta_{0}$ of the layers with respect to the isotherms is 
large, the dynamics of the front must obviously be that of a 
non-faceted crystal as long as the deformation of the front remains 
small, that is, below $V_{c}$ and in a small range of $V$ above $V_c$.  
Facets only appear at higher $V$ when the deformation of the interface 
is large.  A relatively smooth transition from the non-faceted 
to a partly faceted dynamics may then be observed.  This is the 
experimental configuration considered in this study.

In this paper, we study the directional-solidification dynamics of the 
front associated to the nematic-smectic B transition of the liquid 
crystal $\rm{ C_4H_9-(C_6H_{10})_2CN}$ (in short, CCH4).  A long-range 
order exists in the direction perpendicular to the molecular layers in 
the smectic B phase, so that this phase actually is a lamellar 
crystal.  Previous free-growth studies have indeed shown that the 
nematic-smectic B fronts of the $n=3,4,5$ members of the series CCHn 
(where $n$ stands for the number of carbon atoms in the aliphatic chain) 
have a single facet direction parallel to the molecular layers of the 
smectic phase, and are rough in all other directions 
\cite{buto95,tobo96,gora96}.  Moreover, they have no unstable 
orientations in a direction perpendicular to the molecular layers, 
contrary to the Smectic A-Smectic B fronts previously studied in 
directional solidification by Oswald {\it et al} 
\cite{meos90,osme89,meos91}.  The present study is performed in thin 
(12 $\mu m$-thick) samples and in the planar configuration (director 
parallel to the plane of the sample), in order for the front 
--including the facets, if any-- to remain perpendicular to the sample 
plane.  Practically, the system is thus a 2D one.

We shall mostly focus on a type of solitary wave appearing near the 
Mullins-Sekerka threshold, called "faceton" because it contains a 
single small facet traveling along the front at such a velocity that 
the normal growth rate of the facet, i.e., its growth rate with 
respect to the liquid, is generally much smaller than $V$.  Such a 
phenomenon, which has never been observed before, to our best 
knowledge, is obviously highly specific to faceted directional 
solidification, and therefore particularly interesting from our 
present viewpoint.  A preliminary comment about the nematic-smectic B 
facets in the CCHn series is in order.  The growth rate of a facet is 
controlled by the dynamics of the molecular steps flowing along it.  
Therefore it crucially depends on whether, or not, the facet contains, 
or is connected with step sources \cite{buca51,ch89}.  When no step 
source is available, the facet grows through nucleation and spreading 
of new terraces (surface nucleation), which is a very slow process at 
low undercooling.  In fact, the growth rate of a perfect facet is 
totally negligible when the undercooling is lower than some finite 
value.  Such a behavior ("blocked" facets at low undercoolings) has 
clearly been observed during the solidification of many, but not all 
the studied faceted systems. What concerns us here is that it was not 
observed during the free growth of the smectic B phase of CCH3, 
despite the strongly faceted aspect of the growing crystals 
\cite{buto95,tobo96,gora96}.
Numerical simulations in which a cusp-like minimum of $\gamma(\theta)$
but no anisotropy of $\beta$ was taken into account satisfactorily
reproduced the observed growth shapes.
Thus the observation of a facet on a macroscopic scale would
not necessarily mean the presence of a singularity in $\beta$.
In order to clarify this point in the case of CCH4, we 
report, in Section III, preliminary observations in free growth 
showing that the nematic-smectic B facet of CCH4 is capable of 
remaining immobile at undercoolings lower than $0.1 K$.  Thus, 
in CCH4 at least, the nematic-smectic B fronts can form growth facets.

\section{Experimental}
\label{experimental}

The relevant material parameters of the liquid crystal CCH4 (MERCK IS-0558)
can be found in ref.  \cite{tobo96}.
The residual impurities, the chemical nature of which is 
unknown, were characterized as regards solidification by the usual 
methods (see below).  We found that the impurity content at the outset 
of the experiments was reproducible, but slowly increased during the 
experiments, indicating that the product was undergoing a 
decomposition in the nematic phase, as previously noticed and analyzed 
for the case of CCH3 \cite{toeb99}.  The nematic - smectic B 
transition temperature $T_{NS}$ was generally of about $53.1 K$ in 
fresh samples.  Fig.  \ref{tns-t} shows $T_{NS}$ measured as a function 
of time in one sample. It can be seen that the decomposition rate is 
sufficiently slow not to severely perturb a solidification run, but 
sufficiently rapid to prevent us to carry out several successive runs 
with the same sample. Outgasing the as received product resulted in 
a significant slowing down of the decomposition process.

We have studied the crystal structure of the smectic B phase of CCH4 
by low-angle X ray diffraction \cite{le}.  As expected, this phase is 
basically an AB-type stacking of hexagonal layers.  The parameters are 
approximately $a=5.9$\AA \ and $c =29$\AA, which is in accordance with 
the data available for the other members of the homologue series 
\cite{brle81}.  The hexagonal layers however appear to be 
slightly distorted, which may entail the existence of superstructures 
in the layers.


A schematic view of a thin-sample directional-solidification 
experiment is shown in Fig.  \ref{dirsol1}.  A detailed description of 
our setup is given elsewhere \cite{akfa95,akfa98}.  
In this study, the samples were 
made of two parallel glass plates separated by 12 $\mu m$-thick 
plastic spacers.  Their useful width was of 9 mm and their length of 
60 mm.  They were filled under an argon atmosphere at a temperature 
higher than $T_{NS}$, and then cooled down to room temperature.  
Numerous smectic crystals appeared by heterogeneous nucleation during 
cooling.  The samples were placed in the thermal gradient, and a 
smectic B crystal of known orientation (determined through the 
observed value of $\theta_0$) was selected by a method to be explained 
shortly.  The sample was annealed at rest for about 30 minutes in 
order to homogenize the concentration in the liquid.  $V$ was then 
switched to a chosen value, left for given time at this value, and 
then increased step by step.  The temperature gradient at the growth 
front was of 53 $Kcm^{-1}$, unless otherwise mentioned.  The 
pulling velocity was in the range of 0.3 -- 30 $\mu ms^{-1}$.  The 
observations were made with a polarizing microscope (Leica) 
equipped with a CCD camera.  The video signal was  
analyzed with digital image processing.

\vspace*{-0.4cm}
\begin{minipage}{8.3cm}
\begin{figure*}
\epsfxsize=8.5cm
\centerline{\epsfbox{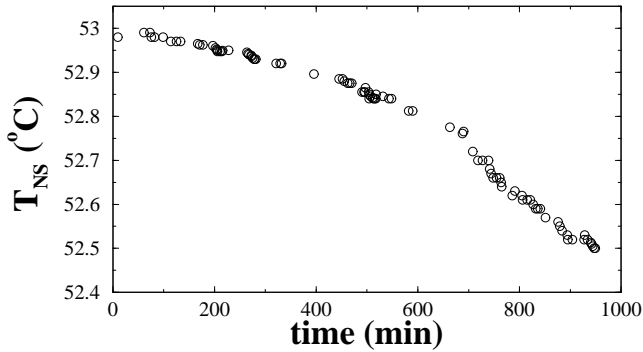}}
\caption{ The nematic - smectic B equilibrium temperature in a CCH4 
sample as a function of time.  $T_{NS}$ was measured by controlling 
the temperature of a free-growth stage in order to keep a small 
smectic B crystal in quasi equilibrium with the nematic.  The relatively 
low initial value of $T_{NS}$ indicates that the sample was rather 
impure at the outset.}
\label{tns-t}
\end{figure*}
\end{minipage}
\smallskip

It is obviously crucial for our experiments that large smectic B 
crystals of arbitrary orientation might be selected.  To this aim, we 
have studied the influence of various treatments of the inner sides of 
the glass plates.  Three types of plates were used : untreated plates, 
plates covered with a mono-oriented thin film of 
poly(tetrafluorethylene) (PTFE) prepared by friction transfer at 
$T\approx 200~^oC$ \cite{dado97}, or with a $\approx$ 100\AA-thick 
layer of Al or In deposited by oblique evaporation.  In the nematic 
phase, the orientation of the director was essentially
planar in all samples.  The director was more or less 
aligned along the direction of friction, or deposition, in treated 
samples, but domains corresponding to small (a few degrees) variations 
in the orientation of the director, still existed (see Fig.  
\ref{recoilfig} below).  This inhomogeneity of the nematic phase 
caused but minor perturbations in our experiments, since the phenomena 
of interest turned out to be essentially independent of the 
orientation of the nematic director.  In all samples, the smectic B 
phase had a planar orientation, but was divided into different 
crystals (or grains) corresponding to a different value of 
$\theta_{0}$.  The surface treatment gave a pronounced preferential 
distribution of $\theta_{0}$ among the various grains, facilitating 
the selection of the desired value of $\theta_{0}$.  The size of the 
selected smectic B grain was increased by a method consisting of 
forcing the crystal to grow through a funnel shaped obstacle 
\cite{akfa95}.  By this method, smectic grains of a millimetric width, 
and arbitrary values of $\theta_0$ were obtained.

\section{Characterization of the system}
\label{prelimresults}

\subsection{Free growth at small undercoolings}
\label{freegrowth}

The observations reported in this Section were performed with a 
free-growth setup similar to the one described in ref.  \cite{boto00}, 
in which the changes in the undercooling are produced, via the 
Clausius-Clapeyron effect, by a sudden pressure change at constant 
temperature, and are therefore quasi instantaneous 
\cite{boto00,coko98}.  The samples were the same as those used in 
directional solidification.  At the beginning of the experiments, the 
samples were heated step by step until only one small smectic B 
crystal was left in the nematic.  The sample was maintained at 
constant temperature until the changes in the shape of the crystal 
became very slow (this took about $20$ minutes).  Admittedly, this 
shape is not the exact equilibrium shape of the crystal, but it 
exhibits clear reproducible features, namely, long facets parallel to 
the smectic layers and rounded ends in the perpendicular direction ( 
Fig.  \ref{freekep} a.), which is enough for our present purpose.  It 
should be noted that the observed near-equilibrium shape clearly shows 
the absence of a forbidden orientation range around $\delta \theta = 
90^{o}$, where $\delta \theta$ is the deviation of the  
interface from the direction of the molecular layers, but suggests that 
the facet might actually be limited by a sharp edge, i.e., the 
interface might be unstable at small values of $\delta \theta$.  The 
fact that we have not observed the Herring instability \cite{he51,osme89} 
in directional solidification at the lowest 
explored value of $\theta_0$ indicates that this forbidden orientation 
range is very small ($< 2^{o}$), if it exists at all.

A sudden increase of the undercooling $\delta T$ was applied at time 
$t~=~0$, and the subsequent growth of the crystal recorded (Figure 
\ref{freekep}).  
The growth process, which is governed by the 
anisotropic interfacial properties and diffusivities, is very 
complicated.  Its study is beyond the scope of this paper.  Here we 
limit ourselves to the following observation: 
the facets of the smectic B crystals remained blocked within 
experimental uncertainty (their growth rate was lower than about $ 
0.01 \mu ms^{-1}$) at undercoolings lower than $0.1~K$  
(Fig.  \ref{freekep}).  At 
higher undercoolings, they generally grow at a measurable rate.  
The apparent threshold undercooling $\Delta T_{nucl}$ for growth by 
surface nucleation of our system is thus larger than $0.1~K$ and
probably not much larger than this value. 
This estimate of $\Delta T_{nucl}$  is small compared to what it is in 
ordinary solid-liquid systems, but this may be explained by the small 
value of $\gamma$ in our system \cite{uh71,uh82}.  It is also 
possible that, in our thin samples, surface nucleation is in fact 
heterogeneous, i.e., takes place preferentially along the line of 
contact with the glass plates.  The nucleation rate would then depend 
on the treatment of the glass plates.
 
 
\begin{minipage}{8.3cm}
\begin{figure*}
\epsfxsize=7cm
\centerline{\epsfbox{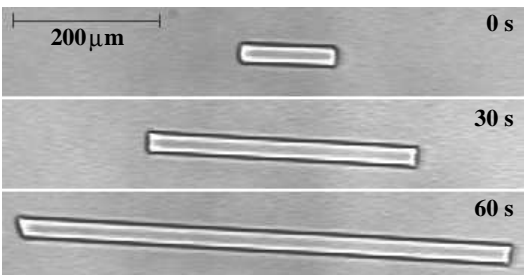}}
\caption{ Free growth.  Successive snapshots of a smectic B crystal of 
CCH4 growing from the nematic phase at $\Delta T = 0.07 K$.}
\label{freekep}
\end{figure*}
\end{minipage}
\smallskip

\subsection{Directional solidification: instability threshold}
\label{dirsol}

The Mullins-Sekerka instability threshold was found to lie between 
approximately 2 and 3 $\mu ms^{-1}$ in all the studied (fresh) 
samples.  No influence of the orientation of the smectic, or the 
nematic was observed within experimental uncertainty.  However, it 
should be noted that this uncertainty was large ($\approx 1~\mu 
ms^{-1}$) for the reason to be explained presently.

Figure \ref{recoilfig} shows a sample at rest, and pulled at a rate 
lower than $V_c$.  Two isothermal fronts are visible, namely, a front 
separating the nematic (N) phase from a smectic B domain (Sm1), and 
at a lower temperature, a front separating Sm1 from a second smectic B 
domain (Sm2).  The nature of the transition from Sm1 to Sm2 is not yet 
clear.  This transition was observed in most, but not all experiments.  
Observations (not reported here) incline us to think that Sm2 is the 
same phase as Sm1, but with a different orientation, thus a different 
interaction energy with the glass plates.  In any case, we need not 
take into account the Sm1-Sm2 front here since this front, when 
present, does not perturb the dynamics of the N-Sm1 front.


It can be seen in Fig.  \ref{recoilfig} that the nematic-smectic B 
front remains planar during solidification at $V < V_{c}$, except for 
small, long-wavelength distortions due to the presence of domains in 
the nematic phase.  These distortions are larger during solidification 
than at rest, and undergo sudden changes each time the front leaves a 
nematic domain for another.  This phenomenon has thus an equilibrium 
as well as a kinetic 
origin.  In our experiments, it plays the role of a relatively strong 
long-wavelength, low-frequency noise, which blurs some of the 
morphological-transition thresholds of the system.  This is the main 
origin of the aforementioned large uncertainty on the measured values 
of $V_{c}$.  However, we can state with certainty that $V_{c}$ was 
higher than $2~\mu ms^{-1}$ since the distortions caused by nematic 
domains, or any other source of perturbation (e.g., dust particles) 
did not amplify below this velocity.
 
\vspace*{-0.2cm}
\begin{minipage}{8.3cm}
\begin{figure*}
\epsfxsize=8.5cm
\centerline{\epsfbox{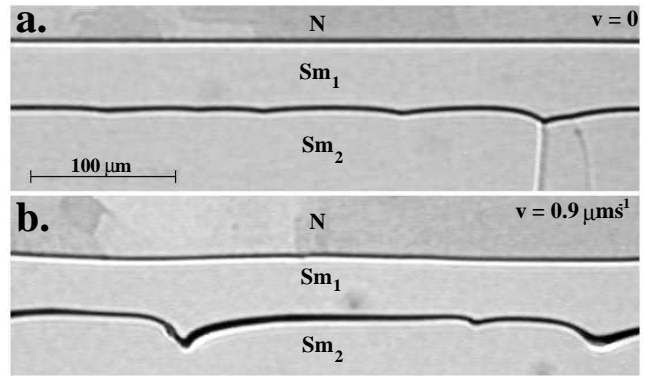}}
\caption{Directional solidification (in this, and all the following 
micrographs, growth is upwards).  N: nematic.  Sm1: smectic B. Sm2: 
smectic B oriented differently from Sm1.  ({\bf a}) Sample at rest 
($V=0$); ({\bf b}) Sample in the process of solidification at $V=0.9 
\mu ms^{-1}$.  Note the domains in the nematic.  Sm1 is a single 
crystal, but Sm2 a polycrystal, as shown by the presence of cusps on 
the Sm1-Sm2 front.}
\label{recoilfig}
\end{figure*}
\end{minipage}
\smallskip

\subsection{Solute redistribution transient}
\label{soluteredist}

When $V$ is smaller than $V_{c}$, the front reaches a stationary 
planar state through the so-called solute redistribution transient.  A 
recoil curve --i.e., the curve representing the variation of the 
position (or temperature) of the planar front as a function of time 
during the initial transient of a particular run-- is reproduced in 
Fig.  \ref{recoil}.  
It is well known that information about the 
relevant properties of the solute (diffusion coefficient $D$ in the 
liquid, partition coefficient $K$, thermal gap $\Delta T_o$) can be 
gained from the characteristics of the transient, and the value of 
$V_c$.  We have utilized this method in order to characterize the 
unknown impurity playing the role of solute in our system.

The threshold velocity, and the amplitude of the solute 
redistribution transient are given by $V_c\approx 
(1+KD^s/D) D G/\Delta T_o $ (${D^s}$: diffusion coefficient in the 
solid) and $\Delta T_o$, respectively \cite{bofa01}.  
By fitting the recoil data using the Warren-Langer
approximate theory \cite{wala93} (Fig. \ref{recoil}) and assuming 
$V_c = 2.5 \mu ms^{-1}$ and $KD^s/D << 1$, we obtained  $K=0.12$. 
This gives $D = 80~\mu m^2s^{-1}$ and $\Delta T_o = 0.2~K$.
These data give us no information about $D^s$, but there is 
good reason to believe that our system is a two-sided one --i.e., that 
$D^s$ is not much smaller than $D$ \cite{meos91}.

\vspace*{-0.4cm}
\begin{minipage}{8.3cm}
\begin{figure*}
\epsfxsize=9cm
\centerline{\epsfbox{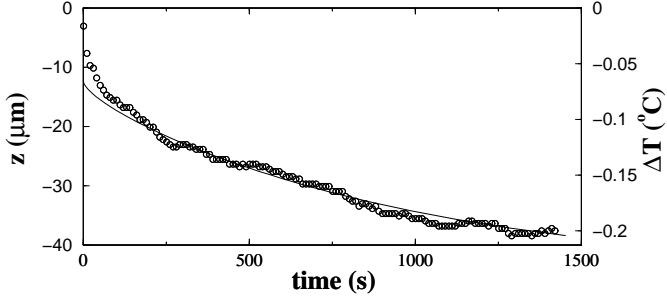}}
\caption{ Recoil curve at $V=0.9~\mu ms^{-1}$.  Same run as in Fig.  
\ref{recoilfig}.  Continuous line:  best fit according to the Warren-Langer
approximation.  
The rapid decrease at the onset of the recoil 
is an instrumental effect.}
\label{recoil}
\end{figure*}
\end{minipage}
\smallskip

\section{Results}
\label{results}

\subsection{Morphology diagram}

A diagram displaying the observed morphologies as a function of the 
pulling velocity and the orientation of the smectic B crystal is shown 
in Fig.  \ref{gamma-v1}.  

\begin{minipage}{8.3cm}
\begin{figure*}[h]
\epsfxsize=8.5cm
\centerline{\epsfbox{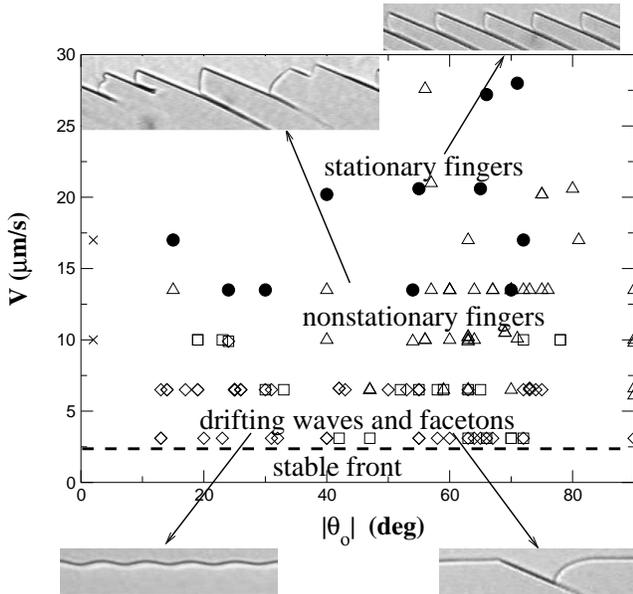}}
\caption{Morphology diagram.  Measurement points: waves and facetons 
($\diamond$), facetons and unstationary faceted fingers ($\Box$), 
unstationary faceted fingers ($\triangle$),  stationary faceted fingers 
($\bullet$) and unstable facets (x).  Heavy 
dashed line: Mullins-Sekerka instability threshold.  
Inset micrographs : see Fig.  \ref{morph}. }
\label{gamma-v1}
\end{figure*}
\end{minipage}
\smallskip

It can be seen that the sequence of 
morphologies observed as a function of $V$ for a fixed value of 
$\theta_0$ is the same for all values of $\theta_0$, except for those 
close to $0^o$ (facets parallel to the growth front) or $ 90^o$ 
(facets perpendicular to the growth front).  This generic sequence 
is illustrated in Fig.  \ref{morph}.

\begin{minipage}{8.3cm}
\begin{figure*}
\epsfxsize=8.5cm
\centerline{\epsfbox{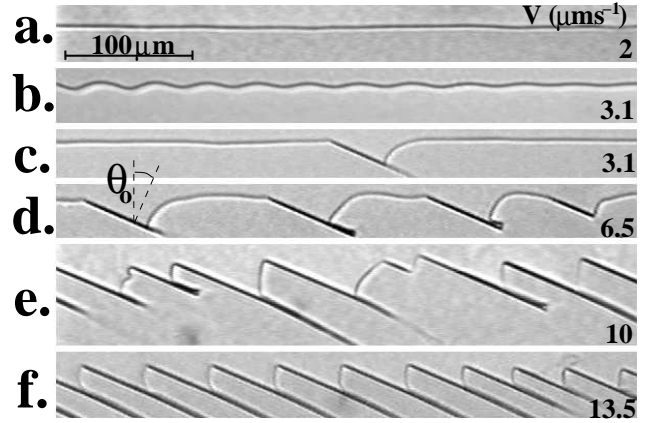}}
\caption{ The different growth morphologies observed as a function of 
$V$ for $\theta_0= 25^o$.  ({\bf a}) Planar front; ({\bf b}) Drifting 
shallow cells; ({\bf c}) Drifting faceton (stationary mode); 
({\bf d})  Drifting facetons (oscillatory mode) at different stages 
of their oscillation cycle; see Fig.  \ref{marc14m11} below; 
({\bf e}) Non-stationary array of faceted fingers ; ({\bf f}) 
Stationary array of faceted fingers.  }
\label{morph}
\end{figure*}
\end{minipage}
\smallskip

Small-amplitude, nearly sinusoidal traveling waves appear near the 
instability threshold (Fig.  \ref{morph}.b), in accordance with 
previous observations in two-sided anisotropic systems \cite{meos90}.  
Such weakly nonlinear waves are commonly called "shallow cells".





We observed drifting shallow cells in a broad range of $V$ around the
threshold ($1~\mu ms^{-1} \leq V \leq 8~\mu ms^{-1}$).  In the same 
range of $V$, we also observed "facetons" (Fig.  \ref{morph}.c).  
These solitary waves can propagate in a 
stationary or an oscillatory way.  They appear when the amplitude of 
the cells is so large that the tilt angle of the front locally reaches 
the value $\theta_0$ corresponding to the facets.  Most generally, 
this occurs under the effect of perturbations due to the nematic domains. 
The frequency of creation of facetons, and thus  
their average number by unit length of the front increases as $V$ 
increases.  When the average spacing of the facetons becomes smaller 
than their width ($\approx 200\mu m$), they cease to behave as 
non-interacting objects.  In fact, they disappear altogether, giving 
way to arrays of much narrower objects, called faceted fingers (Fig.  
\ref{morph}.e).  This occurs at about $ 8\mu ms^{-1}$.  However, this 
transition is strongly noise dependent, and thus relatively 
ill-defined from an experimental viewpoint.  Shallow cells and 
facetons are studied in detail in the next section.



\begin{minipage}{8.3cm}
\begin{figure*}
\epsfxsize=8.5cm
\centerline{\epsfbox{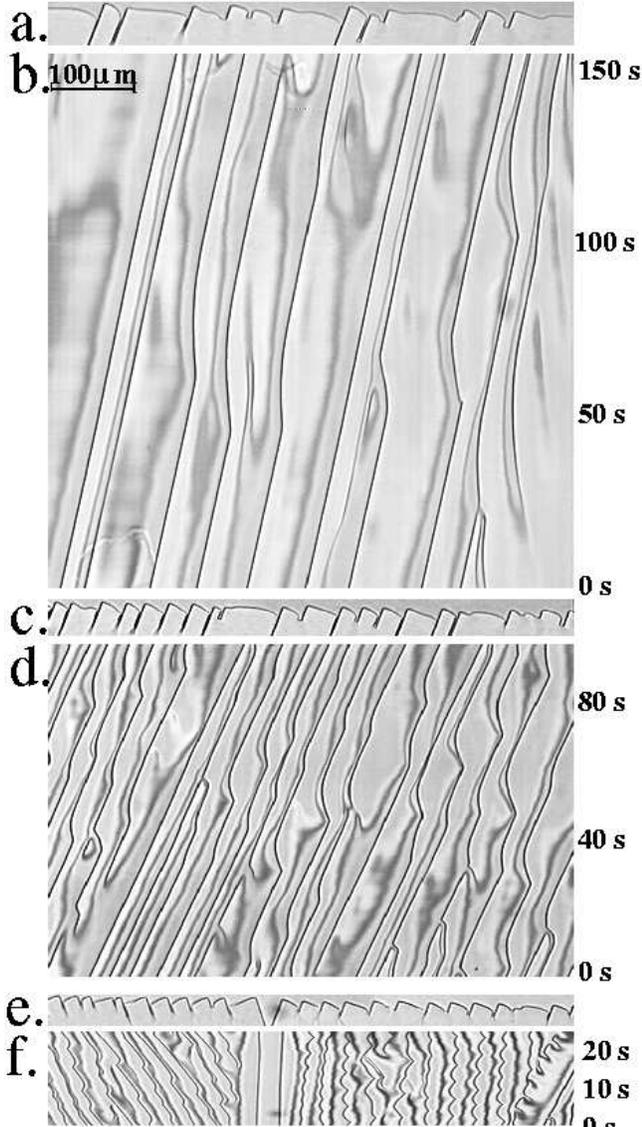}}
\caption{Transition from isolated facetons to faceted fingers for 
$\theta_0=-70^o$ in an aged sample.  ({\bf a}) $V=3.1~\mu ms^{-1}$ 
(snapshot of the front); ({\bf b}) Corresponding spatiotemporal 
diagram (time series of the intensity distribution along a line 
located $20~\mu m$ below the front); ({\bf c}) $V=6.5~\mu ms^{-1}$; 
({\bf d}) Corresponding spatiotemporal diagram; ({\bf e}) $V= 13.5~\mu 
ms^{-1}$.  Note that another grain ($\theta_0=73^o$) appears in the 
leftmost part of the figure; ({\bf f}) Corresponding spatiotemporal 
diagram.}
\label{marc14m23}
\end{figure*}
\end{minipage}
\smallskip

The arrays of faceted fingers, which are observed above $ 8~\mu 
ms^{-1}$ exhibit a relatively sharp transition from an unstationary
(Fig. \ref{morph}.e and Fig. \ref{marc14m23}) to a stationary 
dynamics as $V$ increases
(Figs. \ref{morph}.f and \ref{marc20m14}; the dispersion appearing in 
Fig. \ref{gamma-v1} is mostly due to the aging of the samples).
The spatiotemporal diagrams of 
the unstationary arrays shown in Fig.  \ref{marc14m23} reveal the 
transitory or local existence of well-defined oscillatory modes.  
These modes become more and more apparent as $V$ increases because the 
oscillation period $T_{osc}$ is a rapidly decreasing function of $V$ 
(Fig.  \ref{gammaandv-to}.a).  This strongly suggests the existence of 
a homogeneous oscillatory bifurcation of the high-$V$ stationary 
patterns as $V$ decreases within some narrow range of spacing.

\begin{minipage}{8.3cm}
\begin{figure*}
\epsfxsize=8.5cm 
\centerline{\epsfbox{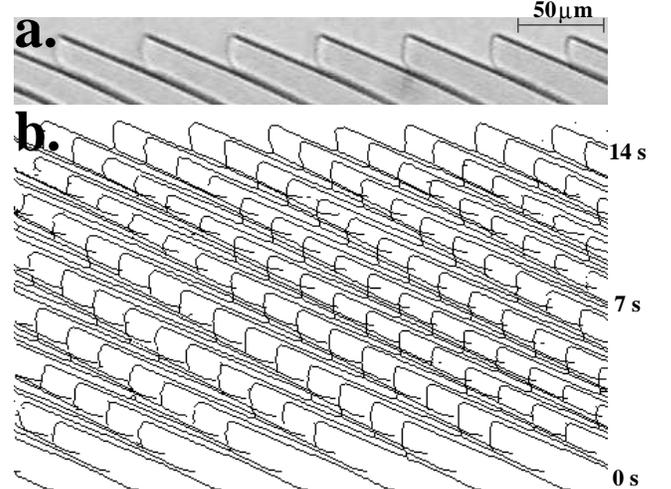}}
\caption{Stationary array of faceted fingers at $V=13.5~\mu ms^{-1}$ and  
$\theta_0=24^o$ ({\bf a.}) Snapshot of the front; ({\bf b.}) 
Spatiotemporal diagram 
(piling-up of skeletonized images of the growth front). } 
\label{marc20m14}
\end{figure*}
\end{minipage}
\smallskip


 
\begin{minipage}{8.3cm}
\begin{figure*}
\epsfxsize=8cm
\centerline{\epsfbox{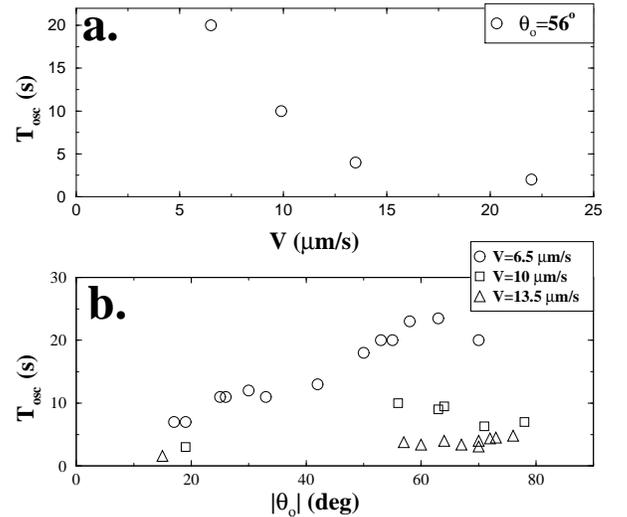}}
\caption{Oscillation period ({\bf a}) as a function of $V$ for 
$\theta_0=56^o$ ({\bf b}) as a function of $|\theta_0|$ for three values of
$V$.  The leftmost point in a) corresponds to an isolated 
oscillatory faceton.  
}
\label{gammaandv-to}
\end{figure*}
\end{minipage}
\smallskip

We now turn to the particular orientations corresponding to the bounds 
of the scanned interval of $\theta_0$. 
When $\theta_0 = 90^o$, the 
system is reflection symmetric.  Shallow cells no longer drift, and 
facetons cease to exist.  The shallow cells break up into 
narrow faceted fingers as $V$ is increased above threshold (Fig.  
\ref{0-90}.a).  The widest faceted fingers, which are the majority 
ones, are not reflection symmetric, whereas the narrowest ones are 
reflection symmetric.  The two opposite but equivalent directions of 
symmetry breaking are equally populated.  The resulting arrays were 
non stationary even at the highest explored values of $V$ (Fig.  
\ref{0-90}.b).  This is very different from what was observed by 
Oswald {\it et al} in smectic A-smectic B fronts for a similar orientation 
of the facet \cite{meos90}.  In that system, because of the existence of 
forbidden directions, the finger tips exhibited pointed triangular 
shapes, and formed stationary arrays.
 
\begin{minipage}{8.3cm}
\begin{figure*}
\epsfxsize=8.4cm
\centerline{\epsfbox{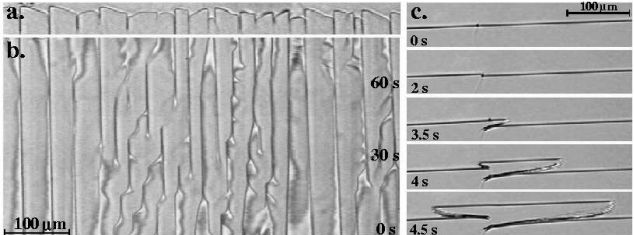}}
\caption{
({\bf a}) Array of symmetry-broken faceted fingers 
at $\theta_0= 90^o$ and $V=10~\mu ms^{-1}$ ; ({\bf b.}) Corresponding 
spatiotemporal diagram;  
({\bf c}) Instability of a facet at $\theta_0= 2^o$ and $V=10~\mu ms^{-1}$. 
}
\label{0-90}
\end{figure*}
\end{minipage}
\smallskip

When $\theta_0$ is sufficiently close to zero, the growth front of 
smectic B grains is entirely occupied by a facet at any value of $V$.  
This can be considered as a finite-size effect resulting from the 
following fact: facets are always present in the grooves attached to 
grain boundaries for whatever values of $\theta_0$ and $V$; the 
stationary size of these facets is more or less proportional to 
$1/(tan |\theta_0|)$; they thus occupy the whole grain when $|\theta_0|$ is 
lower than a certain value, which is of about $2^o$ for a grain size 
of $500 \mu m$.  At sufficiently high $V$, these long facets break up 
through the mechanism illustrated in Fig.  \ref{0-90}.c.  It is not 
necessary to repeat here the description of this process, which has been 
presented by other authors \cite{meos91}.  We simply note that, in our 
fresh samples, this instability was observed to result from the 
occasional collisions of the front with defects (domain walls, dust 
particles) present in the nematic.  In the less pure samples, it was 
superseded by another well-known process, namely, the nucleation of 
crystals in the undercooled melt ahead of the front \cite{meos90}.  
Both mechanisms give rise to more or less permanently cyclic growth regimes.


\subsection{Near-threshold patterns}
\subsubsection{Drifting shallow cells}

Most generally shallow cells appeared in the form of a noise-induced wave 
packet. A spontaneous homogeneous growth of the 
cells was never observed with certainty.  We attribute this fact to 
the interplay between shallow cells and facetons (see below).  At, or 
below $2 \mu ms^{-1}$, noise-induced wave packets systematically 
disappeared when the source of noise disappeared, as already 
mentioned.  At higher $V$, they evolved as illustrated in Figs.  
\ref{wavesspatio} and \ref{m26seq}.


\begin{minipage}{8.3cm}
\begin{figure*}
\epsfxsize=8.3cm 
\centerline{\epsfbox{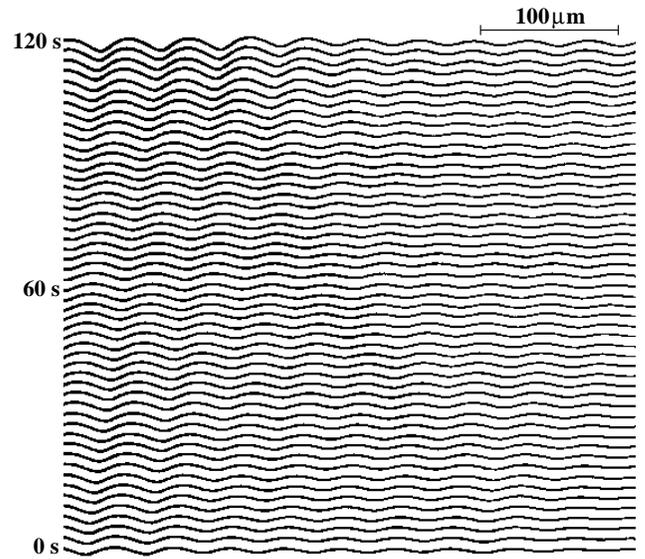}}
\caption{ Spatiotemporal diagram of a drifting wave packet; 
$ V=3.1~\mu ms^{-1}$,  $\theta_0=25^o$. }
\label{wavesspatio}
\end{figure*}
\end{minipage}
\smallskip

A careful analysis of the spatiotemporal diagram of Fig.  
\ref{wavesspatio} has shown that (i) the cells are initially 
sinusoidal; (ii) they grow in amplitude with a uniform amplification 
rate of $\approx 0.002 s^{-1}$; (iii) the amplest cells are no longer 
sinusoidal at the end of the time sequence, (iv) the spacing 
$\lambda$ and the drift velocity $V_{d}$ are uniform in space and 
constant in time.  $V_{d}$ is thus amplitude independent.  This is in 
keeping with the idea that this sequence is the initial stage of the 
usual amplification process leading from a linearly unstable state to 
a stationary weakly nonlinear regime.  The final regime was not observed 
because the process was 
interrupted by an external perturbation giving rise to a faceton.



The traces on the lefthand side of Figure \ref{m26seq} are the 
trajectories of three oscillatory facetons.  These objects are studied 
below.  For now, the point of interest is that the rearmost faceton 
leaves behind a region of the front which is free of detectable 
shallow cells (see also Figs.  \ref{marc20m11} and \ref{marc14m11} 
below).  The cells reappear at $\approx 200~\mu m$ from the faceton, 
and then amplify following a process entirely similar to the above 
one, except for two points: (i) in the present case, the amplification 
rate ($\approx 0.02 s^{-1}$) is much larger than in the preceding 
case, since $V$ is higher, and (ii) a 
stationary regime of nonlinear shallow cells is reached.  This 
confirms clearly, although only semi-quantitatively, that the system 
admits stationary weakly nonlinear cellular states within a measurable 
range of $V$ above $V_{c}$.  These states are metastable with respect 
to the formation of facetons.  Also we note that the direction of 
drift of the cells is opposite to that of facetons.  This is somewhat 
of a surprise since, in other systems, shallow cells and facets have 
been found to drift in the same direction \cite{meos90}.

\begin{minipage}{8.3cm}
\begin{figure*}
\epsfxsize=8.3cm 
\centerline{\epsfbox{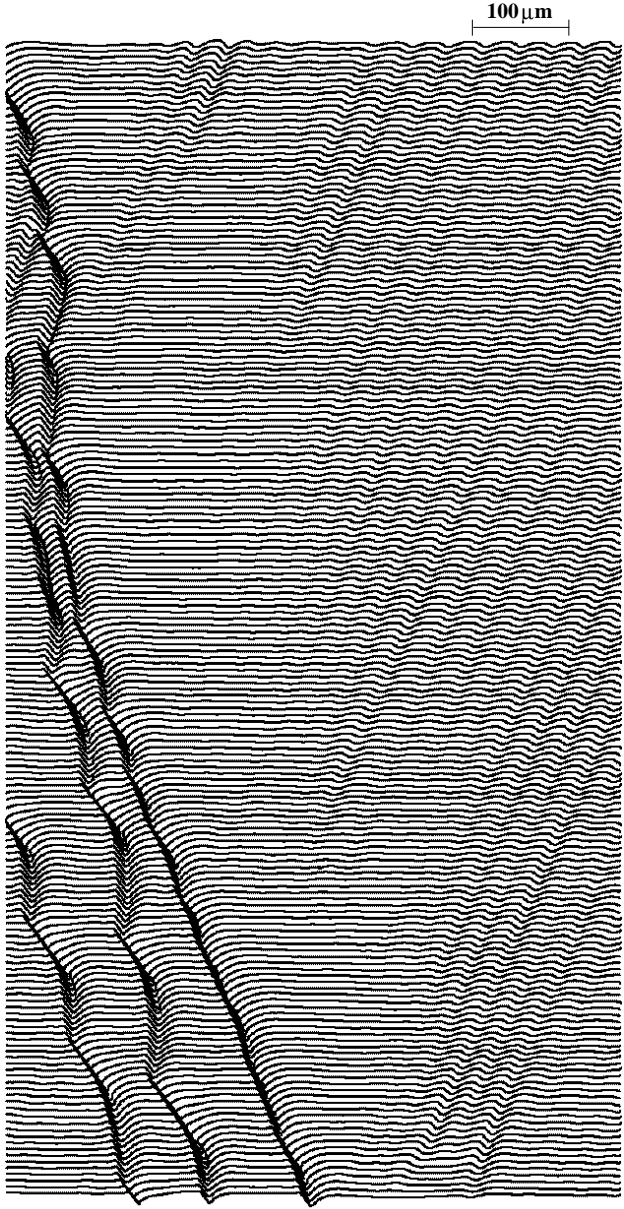}}
\caption{Spatiotemporal diagram.  The three traces on the lefthand 
side are the trajectories of oscillatory facetons drifting leftwards.  
Note the disappearance of the cells (which drift rightwards) in the 
wake of the rearmost faceton.  A temporary exception to this rule is visible 
near the end of the recording, when the faceton emits a packet of 3 or 
4 cells.  This exception is only apparent, however, since this occurs 
during a period of time when the faceton no longer exists (it is 
drifting rightwards).  $V=6.5~\mu ms^{-1}$,  $\theta_0=55 ^o$,  
recording time = 250 s.  }
\label{m26seq}
\end{figure*}
\end{minipage}
\smallskip


The measured values of $|V_{d}|$ and $\lambda$ are plotted in Fig.  
\ref{wav} as a function of $\theta_0$ for a given value of $V$.  
The data are compatible with the fact that
$V_d(\theta_0)$ must go to zero at $\theta_0=0^o$ and $90^o$ 
for symmetry reasons.  The maximum is at about $70^o$, and 
corresponds to a relatively large value of $V_d/V$, indicating 
that the system is strongly anisotropic even in the orientation range 
in which the interface is rough.  

\begin{minipage}{8.3cm}
\begin{figure*}
\epsfxsize=8.3cm 
\centerline{\epsfbox{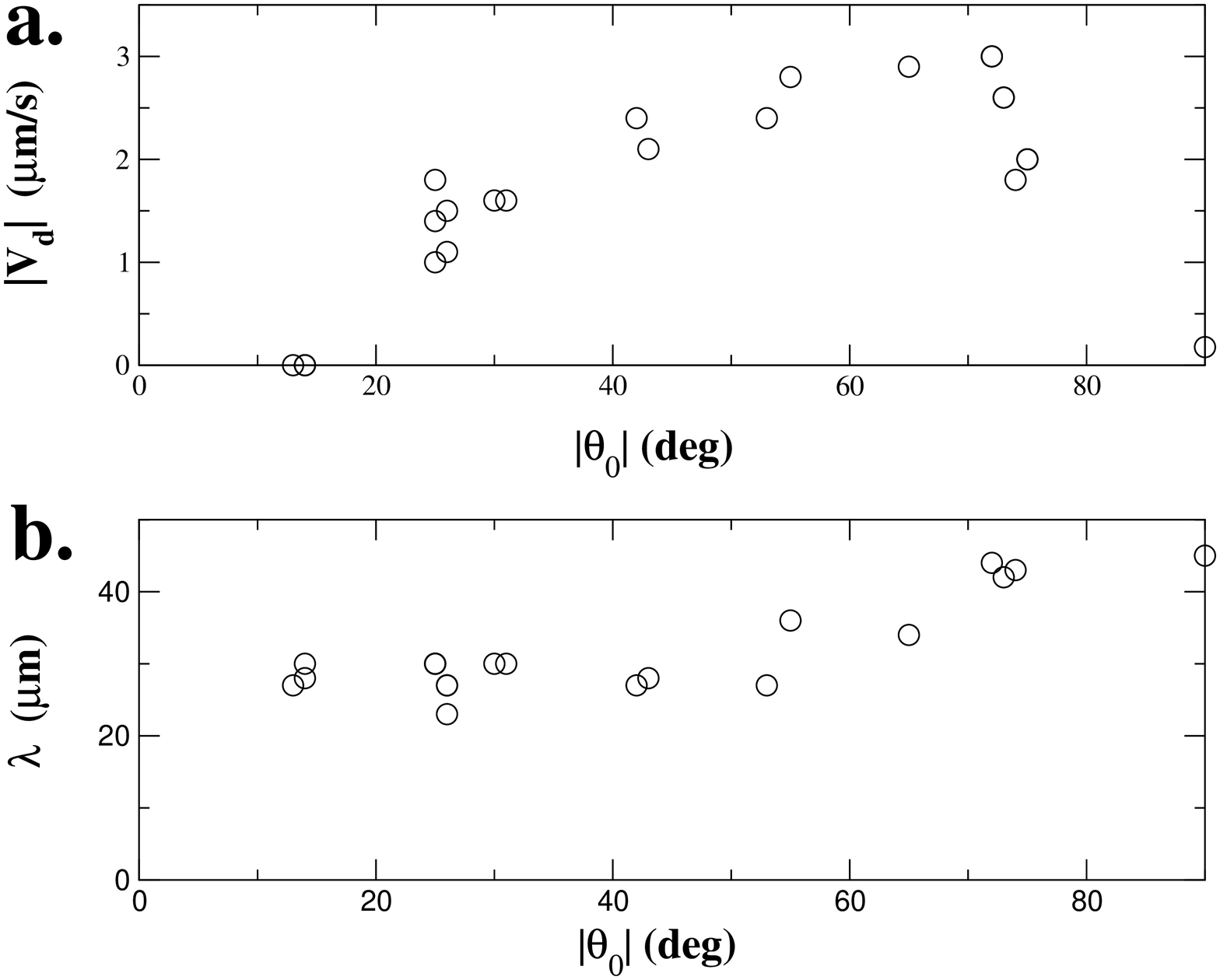}}
\caption{Drift velocity ({\bf a}) and wavelength ({\bf b}) of the 
cells as a function of the tilt angle of the facet at $V=6.5 \mu 
ms^{-1}$.  }
\label{wav}
\end{figure*}
\end{minipage}
\smallskip

We have noted above that $V_{d}$ 
seems to be independent of the amplitude of the cells.  It is thus 
legitimate to admit (but not certain) that the measured value of 
$V_{d}$ is the same as in the linear regime.  We have performed a 
linear stability analysis of the planar front of a two-sided system 
taking into account the anisotropy of the diffusion in the two bulk phases 
(nematic and smectic B), and that of the linear kinetic 
coefficient $\beta$ (the anisotropy of $\gamma$ does not come into 
play in a linear calculation \cite{cose76}).  We have solved the 
dispersion equation numerically under various assumptions concerning 
the orientation dependences of $D$, $D_{s}$, and $\beta$, which are 
not known.  Qualitatively, the results can be summed up as follows 
\cite{bofa01}.  We find that the observed sign and absolute value of 
$V_{d}(\theta_0)$ could be ascribed to diffusion anisotropy only if, 
in the smectic B phase, the impurities diffused much faster through 
the smectic layers than parallel to them, which is very unlikely to be 
true.  Thus the observed drift of the shallow cells is most probably 
due to kinetic anisotropy.  In such a case, the sign of $V_{d}$ is 
given by $-d\beta/d\theta$ \cite{cose76}. In conclusion, the 
observed direction of drift of the shallow cells (if it is really the 
same as in the linear regime) indicates that, in our system, $\beta$ 
increases as $\delta \theta$ increases.  This result poses no 
particular problem except for the vicinal domain, in which $\beta$ is 
expected to be more or less proportional to the reciprocal of the step 
density, and hence to the reciprocal of $\delta \theta$ \cite{chco93}.  
The cross-over from the vicinal to the rough domains as $\delta 
\theta$ increases should thus manifest itself through a change in the 
sign of $V_{d}$.  It is tempting to assume that this cross-over 
corresponds to the zero of $V_d(\theta_0)$ which perhaps appears near 
$12^o$ in Fig. \ref{wav}.a.  However, the observation of 
macroscopic facets drifting in the same direction as the shallow cells 
disproves this assumption, and indicates that the vicinal domain is 
actually very narrow in our system (see below).

\subsubsection{Stationary facetons}

The spatiotemporal diagram of a stationary faceton is shown in Fig.  
\ref{marc20m11}.  Clearly, a faceton is a solitary wave consisting of 
a macroscopic facet and a broad rounded finger separated from each 
other by a very thin liquid groove.  The regularity of the 
spatiotemporal diagram shows that facetons, once formed, are quite 
stable.  In particular, they absorb the shallow cells which they may 
encounter ahead of themselves without being modified, and seem to be 
insensitive to the perturbations caused by the nematic domains.  The 
depth of the facet --i.e., the distance $\Delta z _f$ between the two 
edges of the facet along the z axis-- is difficult to measure with 
accuracy because the lower edge, located near the bottom of the 
groove, is generally not resolved.  However, it is certain that 
$\Delta z _f$ is in the $30-50 \mu m$ range 
(the difference of temperature $\Delta T _f$ between the two edges 
is thus in the range $0.15-0.25~K$), 
and decreases as $\theta_0$ increases.  The upper edge of the facet 
corresponds to a small pointed maximum of the front shape, but it is 
not possible to decide whether, or not, this edge is sharp on a 
molecular scale.  The width of the rounded finger --i.e., the 
extension of the deformed region of the front behind the finger tip-- 
is of about $200~\mu m$.  As mentioned, shallow cells do not develop 
in this region of the front.  The trajectory of the faceton makes a 
small angle with the direction of the macroscopic facet, indicating 
that the normal growth rate of the facet is small but finite. Thus the 
facet is not blocked, and the 
question arises as to its microscopic growth mechanisms.

\begin{minipage}{8.3cm}
\begin{figure*}
\epsfxsize=8.5cm 
\centerline{\epsfbox{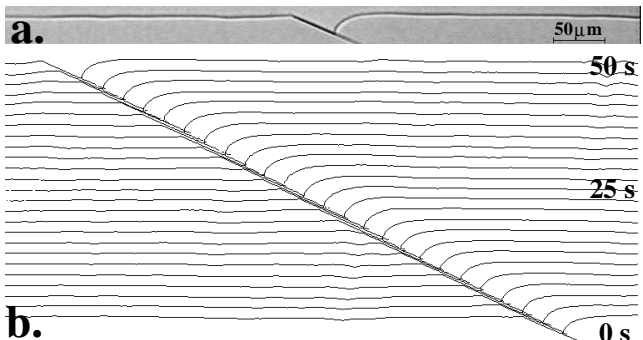}}
\caption{Stationary faceton.  $\theta_0=25^o$, $V=6.5~\mu ms^{-1}$.  
({\bf a.}) Snapshot of the front.  The faint dark line appearing in 
the solid in the continuation of the facet is a thin liquid groove; 
see Fig.  \ref{form}.  ({\bf b.}) Spatiotemporal diagram.  The normal 
growth rate of the facet is $V_{n}~\approx 0.9~\mu ms^{-1} $}
\label{marc20m11}
\end{figure*}
\end{minipage}
\smallskip

Fig. \ref{Vn} displays a large number of values of the normal velocity
of facets $V_{n}$ measured 
in isolated facetons as well as in arrays of faceted fingers for 
various values of $V$ and $\theta_{0}$.  In spite of a large 
dispersion of the data, it is clear that $V_{n}$ is essentially a 
non-zero quantity  
which decreases as $\theta_{0}$ increases, and increases as $V$ 
increases.  The regularity of the stationary facetons or arrays (see 
Figs.  \ref{marc20m11} and \ref{marc20m14}), and the fact that $V_{n}$ 
is very close to zero when $\theta_{0}$ is large allow us to exclude 
screw dislocation growth as the dominant mechanism. Moreover, 
the fact that both $V_{n}$ and $\Delta z _f$ are decreasing functions 
of $\theta_0$ suggests that $V_{n}$ is essentially determined by 
events occurring near the lower edge of the facet.  One may imagine 
either that surface nucleation takes place at a relatively high rate 
at this point, or that the facet is supplied with steps coming from the 
bottom of the groove where the interface is necessarily rough.  In both 
cases, $V_{n}$ would be very sensitive to the details of the 
conformation of the interface in this region.  These details may 
depend on the treatment of the glass plates, which could explain the 
dispersion between values measured in different samples.



\begin{minipage}{8.3cm}
\begin{figure*}
\epsfxsize=8.3cm 
\centerline{\epsfbox{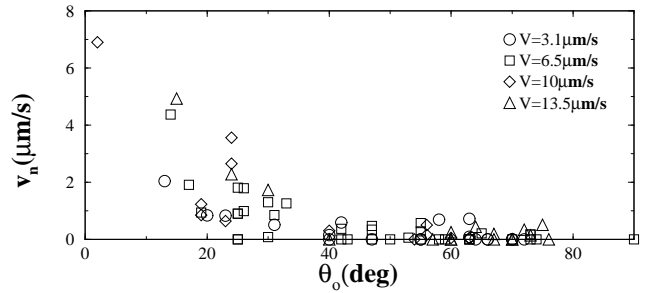}}
\caption{Normal growth velocity of facets belonging to facetons or 
arrays of faceted fingers as a function of the tilt angle of the facet 
for the indicated values of the pulling velocity. In the case of oscillatory
facetons the minimum value of $V_n$ has been plotted. The data point at 
$\theta_0=2^o$ corresponds to the facet shown in Fig. \ref{0-90}.c prior 
to its destabilization.}
\label{Vn}
\end{figure*}
\end{minipage}
\smallskip

\subsubsection{Oscillating facetons}

Fig.  \ref{form} shows a process of formation of facetons in response 
to a perturbation.  Macroscopic facets progressively develop on one 
side of the shallow cells as the amplitude of the latter increases.  
These facets first drift with the same velocity as the shallow cells, 
and then change their direction of drift.  This change is not 
accompanied by any modification in the orientation of the facets within 
experimental uncertainty ($\approx 0.5^o$).  Thus the same macroscopic 
facet can be in two different microscopic states, or growth regimes.  
One of these (the "slow" regime) is that of the stationary state, 
discussed in the preceding section, while the other (the "rapid" 
regime) corresponds to a rough interface.  As announced, we are thus 
led to assume that the cross-over from vicinal to rough interfaces 
occurs at values of $\delta \theta$ lower than $\approx 0.5^o$ in our 
system.  This is indeed surprising since this disorientation 
corresponds to a very low density of steps (less than $1$ per 
$\mu m$), but not impossible.  We also note that the persistence of a 
macroscopic facet while the interface is rough on a microscopic scale 
is explainable by the sole singularity of the $\gamma$ -plot 
\cite{gora96}.


\begin{minipage}{8.3cm}
\begin{figure*}
\epsfxsize=8cm
\centerline{\epsfbox{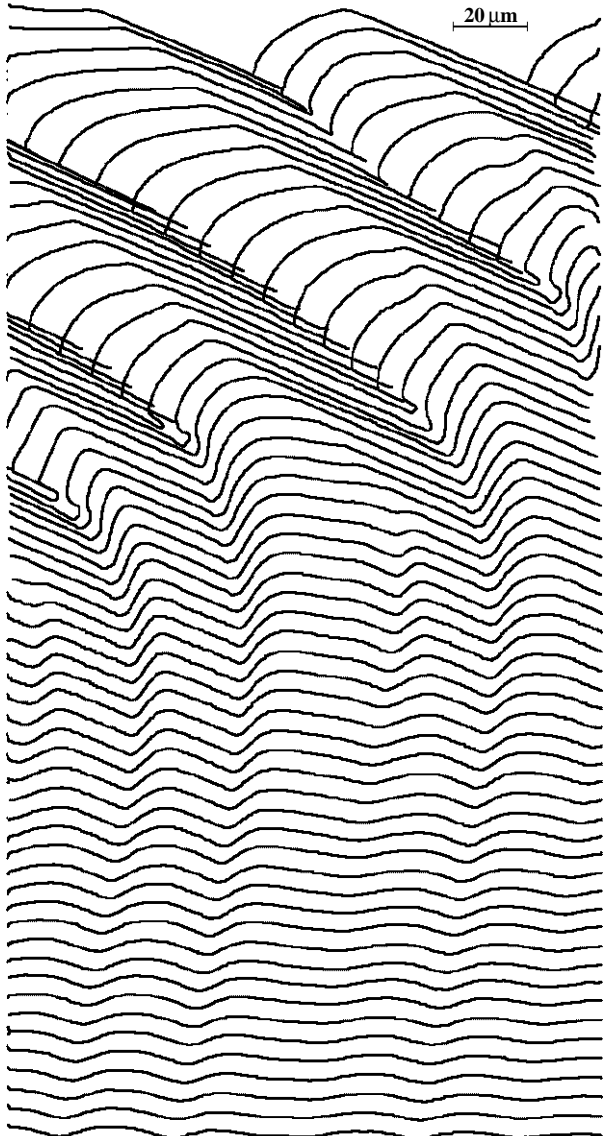}}
\caption{ 
Facetons appearing in response to a perturbation.
Spatiotemporal diagram.  $V=6.5~\mu ms^{-1} $,  
$\theta_0=25^o$,  recording time: 60 s.  Note the opposite signs of 
the drift velocities of the cells and the facets.}
\label{form}
\end{figure*}
\end{minipage}
\smallskip

The existence of two different growth regimes of a macroscopic facet 
is confirmed by the fact that facetons most often adopt an oscillatory 
mode of propagation (Fig.  \ref{marc14m11}).  Obviously, this 
oscillation consists of a more or less ample cycle between the 
aforementioned rapid and slow regimes.  The conditions under which 
facetons are stationary, or oscillatory could not be determined.  In 
fact, stationary facetons were observed much less frequently than, and 
always in coexistence with oscillating facetons.  Moreover, some 
oscillating facetons were regular (Fig.  \ref{marc14m11}), but most of 
them were irregular (Figs.  \ref{m26seq} or \ref{smallfacet}).  It is 
possible that the system intrinsically admits stationary, periodic, 
and more or less chaotic facetons.  However, the following explanation 
is also possible.
 
\begin{minipage}{8.3cm}
\begin{figure*}
\epsfxsize=8.5cm 
\centerline{\epsfbox{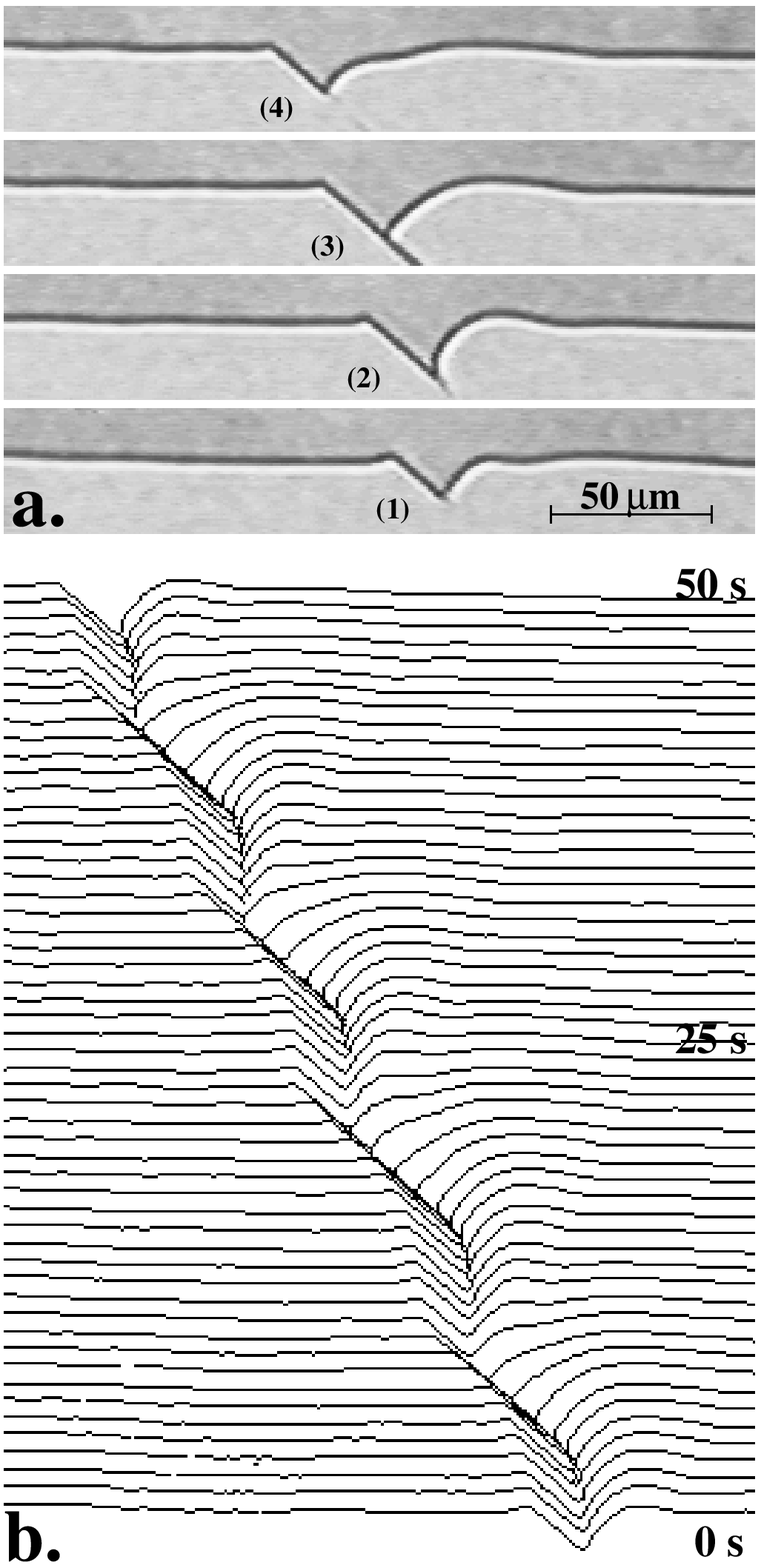}}
\caption{Oscillating faceton.  $\theta_0=42^o$,  $V=6.5~\mu 
ms^{-1}$.  ({\bf a.}) Snapshots of the front at different stages of an 
oscillation period.  ({\bf b.}) Spatiotemporal diagram.  }
\label{marc14m11}
\end{figure*}
\end{minipage}
\smallskip

A careful inspection of Fig. \ref{marc14m11} reveals that the 
transition of the oscillating facetons from a slow to a rapid regime 
corresponds to a sudden pinching-off of the liquid groove, whereas the 
reverse transition from a rapid to a slow regime consists of a 
progressive deepening of the groove.  If we focus on the sole groove, 
this behavior is strongly reminiscent of the periodic pinching-off 
(called cusp instability) of the intercell grooves in nonfaceted 
cellular fronts \cite{kuch89}.  This instability, we recall, is 
most probably of a capillary origin (Rayleigh instability) \cite{br89}, 
and very sensitive to the lattice defects which, in the 
non-faceted systems, are often attached to the groove --in fact, the 
grooves to which subboundaries (low-angle grain boundaries) are 
attached are not subject to the cusp instability \cite{boak01}.
If, by analogy, we 
assume that the intercell groove of facetons, like that of nonfaceted 
cells, is intrinsically subject to an oscillatory Rayleigh 
instability, we are led to the conclusion that the transition of the 
facet from a slow to a rapid regime is a secondary effect due to 
changes occurring in the configuration of the interface near the lower 
edge of the facet.  The presence of lattice defects (e.g., 
subboundaries) emerging into the liquid at the bottom of the groove 
may hinder these changes, suppressing the oscillation.  This would 
explain that facetons are much more often oscillatory than stationary.


\subsubsection{Lattice defects}

Some lattice defects (mostly, grain boundaries) can be detected with 
the optical microscope thanks to the fact that they create macroscopic 
depressions (grooves) of the growth front around the point at which they 
emerge into the liquid.  In our system, these grooves must be partly 
faceted during solidification.  We have lowered the applied thermal 
gradient in some experiments in order to facilitate the observation of 
such grooves.  This allowed us to reveal that the growth front of our 
system is often swept by very small facets, called micro-facets, 
certainly attached to lattice defects emerging into the liquid. 
 
\begin{minipage}{8.3cm}
\begin{figure*}
\epsfxsize=8.3cm
\centerline{\epsfbox{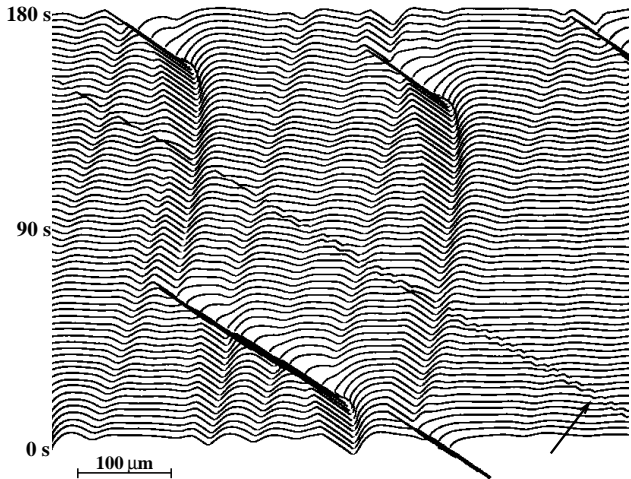}}
\caption{ Spatiotemporal diagram showing shallow cells, oscillating 
faceted solitary waves, and micro-facets (arrow),  $\theta_0=36^o$,  
$V=3.1~\mu ms^{-1}$ and  $G=25~Kcm^{-1}$. }
\label{smallfacet}
\end{figure*}
\end{minipage}
\smallskip

We observed several types of micro-facets, corresponding probably to 
different types of lattice defects.  The micro-facets of the type 
shown in Figure \ref{smallfacet} were relatively easy to identify 
because they travel at a perfectly constant velocity, catching up, and 
running through all the other structures of the front, in particular, 
facetons.  Their drift velocity has thus most probably the maximum 
possible value, i.e., the value corresponding to totally blocked 
facets.  They must be attached to lattice defects 
--stacking faults, or twist subboundaries-- strongly locked onto the 
lamella plane of the smectic.  However, these micro-facets seem to 
have but little effect on the dynamics of the front.  They indeed 
provoke an instantaneous slowing-down of the macroscopic "rapid" 
facets when they collide with them (see Fig.  \ref{smallfacet}), but 
do not trigger a durable transition to the slow regime.  So this 
observation, whatever its intrinsic interest may be, does not cast 
light on the question of the possible role played by lattice defects 
in the dynamics of the facetons.



\section{Discussion}
\label{summary}

We have shown that the directional solidification of a nematic-smectic B 
front in the planar configuration gives rise to a wealth of 
interesting nonlinear phenomena, the most striking of which are the 
stationary or oscillatory "facetons" encountered in the vicinity of 
the Mullins-Sekerka threshold.  These observations raise numerous 
unsolved problems concerning the microscopic growth mechanisms of the 
facets, as well as the nonlinear dynamics of the observed macroscopic 
patterns.  An important question is whether these phenomena are 
specific of the nematic-smectic B fronts, or are of frequent 
occurrence in partly faceted fronts.  In order to clarify this point, 
we are currently searching for similar phenomena in more conventional, 
partly faceted solidification fronts.  Also numerical simulations 
based on a phase-field method are in progress in order to test the 
consistency of the numerous conjectures which we have been led to make 
in order to explain the peculiar dynamical features of the facetons.

\acknowledgments{ The authors wish to thank A-M. Levelut for the help 
in characterizing CCH4 with X-ray diffraction, T. T\'oth-Katona and 
\'A. Buka for many useful discussions and A. Fleury and C. Picard
for their technical assistance. We are also much grateful to 
MERCK (Darmstadt) for kindly providing us with CCH4. This 
research has been supported by a Marie Curie Fellowship of the 
European Community programme IMPROVING HUMAN POTENTIAL under contract 
number HPMF-CT-1999-00132.  
}

\end{multicols}


\vspace*{-0.2cm}

\begin{references}
\bibliographystyle{unsrt}

\bibitem{go92} Solids Far from Equilibrium, 
C. Godr\`eche ed. (Cambridge University Press, 1992).

\bibitem{crho93}
M.~Cross and P.~Hohenberg.
\newblock {\em Rev. of Mod. Phys.}, {\bf 65}, 851 (1993).

\bibitem{la80}
J.~Langer.
\newblock {\em Rev. of Modern Physics}, {\bf 52}, 1 (1980).

\bibitem{cose76}
S. R. Coriell and R. F. Sekerka J. Cryst. Growth {\bf 34}, 157 (1976).

\bibitem{koka96}
P. Kopczynski, W.-J. Rappel and A. Karma, Phys. Rev. Lett., {\bf 77} 3387
(1996).

\bibitem{meos90} F. Melo and P. Oswald, Phys. Rev. Lett., 
{\bf 64}, 1381 (1990).

\bibitem{akfa95}
S. Akamatsu, G. Faivre and T. Ihle, Phys. Rev. E {\bf 51}, 4751 (1995).

\bibitem{akfa98}
S. Akamatsu and G. Faivre, Phys. Rev. E {\bf 58}, 3302 (1998).


\bibitem{shhu91}
D.K. Shangguan and J.D. Hunt, Metall. Trans. A {\bf 22A}, 941 (1991).

\bibitem{fatr97}
L.M. Fabietti, and R. Trivedi, J. Cryst. Growth {\bf 182}, 185 (1997).

\bibitem{osme89}
P. Oswald, F. Melo and C. Germain,
J. Phys. France, {\bf 50}, 3527 (1989).

\bibitem{meos91} F. Melo and P. Oswald, J. Phys. II, {\bf 1}, 353 (1991).

\bibitem{buca51}
W. K. Burton, N. Cabrera and F. C. Frank, 
Phil. Roy. Soc. London {\bf 243}, 299 (1951).

\bibitem{buto95} \'A. Buka, T. T\'oth-Katona and L. Kramer,
Phys. Rev. E {\bf 51}, 571 (1995).

\bibitem{tobo96} T. T\'oth-Katona, T. B\"orzs\"onyi, Z. V\'aradi,
J. Szabon, \'A. Buka, R. Gonz\'alez-Cinca, L. Ramirez-Piscina,
J. Casademunt, and A. Hern\'andez-Machado, Phys. Rev. E {\bf
54}, 1574 (1996).

\bibitem{gora96}
R.~Gonz\'alez-Cinca, L.~Ramirez-Piscina, J.~Casademunt, A.~Hern\'andez-Machado,
  T.~T\'oth-Katona, T.~B\"orzs\"onyi, and {\'A}.~Buka.
Physica D, {\bf 99}, 359 (1996).

\bibitem{ch89}
A. A. Chernov, Contemporary Physics, {\bf 30} 251 (1989). 

\bibitem{toeb99} T. T\'oth-Katona, N. \'Eber and \'A. Buka,
Mol. Cryst. Liq. Cryst. {\bf 328}, 467 (1999) 

\bibitem{le} A.M. Levelut, unpublished results.

\bibitem{brle81} R. Brownsey and A. Leadbetter,
J. Phys. Lett., {\bf 42}, 135 (1981).

\bibitem{dado97} P. Damman, M. Dosi\`ere, M. Brunel and J.C. Wittmann, 
J. Am. Chem. Soc.  {\bf 119}, 4633 (1997)

\bibitem{boto00} T. B\"orzs\"onyi, T. T\'oth-Katona, \'A. Buka
and L. Gr\'an\'asy, Phys. Rev. E {\bf 62}, 7817 (2000).

\bibitem{coko98} J. C. La Combe, M. B. Koss, L. A. Tennenhouse, E. A.
Winsa, and M. E. Glicksman, J. Cryst. Growth {\bf 194}, 143 (1998).

\bibitem{he51} C. Herring, Phys. Rev. {\bf 82} 87 (1951).

\bibitem{uh71} D. R. Ulhmann in "Advances in Nucleation and Crystallization 
in Glasses" (1971) The American Ceramic Society, Columbus, Ohio. 

\bibitem{uh82} D. R. Ulhmann in "Nucleation and Crystallization 
in Glasses" (1982) The American Ceramic Society, Columbus, Ohio. 

\bibitem{bofa01} T. B\"orzs\"onyi, S. Akamatsu and G. Faivre, 
unpublished results.

\bibitem{wala93} J.A. Warren and J.S. Langer, Phys. Rev. E {\bf 47}, 2702 
(1993).

\bibitem{chco93} A. A. Chernov, S. R. Coriell and B. T. Murray, 
J. Cryst. Growth {\bf 132} 405 (1993).

\bibitem{kuch89} P. Kurowski, S. de Cheveign\'e, G. Faivre and C. Guthmann,
J. Phys (France) {\bf 50} 3007 (1989).  

\bibitem{br89} K. Brattkus, J. Phys (France) {\bf 50} 2999 (1989).  

\bibitem{boak01} S. Bottin-Rousseau, S. Akamatsu and G. Faivre, 
unpublished results.



\end{references}
\end{document}